\documentclass[prd, preprint, longbibliography, 12pt]{revtex4-1}

\usepackage{commath}
\usepackage{amsmath}
\usepackage{amssymb}
\usepackage{setspace}
\usepackage{graphicx}
\usepackage{natbib}
\usepackage{float}
\usepackage{tensor}
\usepackage[utf8]{inputenc}
\usepackage{amsfonts}
\usepackage{braket}
\usepackage{esint}
\usepackage{breqn}
\usepackage{IEEEtrantools}
\usepackage{hyperref}

\usepackage[english]{babel}
\usepackage{graphicx}
\usepackage{amsmath}

\usepackage{amssymb}
\usepackage{multirow}
\usepackage{url}
\usepackage{tikz}
\usepackage{braket}
\usetikzlibrary{decorations.markings}
\usepackage{subcaption}
\usepackage{cancel}
\usepackage{mathtools}
\usepackage[utf8]{inputenc}
\usepackage{xargs, ifthen}
\usepackage[breakwords]{truncate}

\usepackage{hyperref}
\DeclareMathAlphabet{\pazocal}{OMS}{zplm}{m}{n}

\allowdisplaybreaks
\def\s[#1\s]{\begin{align}\begin{split}#1\end{split}\end{align}}
\def\[#1\]{\begin{align}#1\end{align}}
\begin{document}
 
 % ! TEX spellcheck
 %
%\ \vskip 1.0 in

\begin{center}
{ \large \bf Trace dynamics, and a ground state in }\\
{ \large{\bf  spontaneous quantum gravity} }
% Submitted to arXiv hep-th on August 2, 2020

\vskip 0.2 in

{\large{\bf Abhinash Kumar Roy$^{a1}$, Anmol Sahu$^{b2}$ and Tejinder P.  Singh$^{c3}$ }}

\medskip

{\it $^a$Indian Institute of Science Education and Research Kolkata 741246, India}\\

{\it $^b$Indian Institute of Science Education and Research, Pune 411008, India}\\

{\it $^c$Tata Institute of Fundamental Research, Homi Bhabha Road, Mumbai 400005, India}\\

\bigskip

 \; {$^{1}$\tt akr16ms137@iiserkol.ac.in}, {$^{2}$\tt anmol.sahu@students.iiserpune.ac.in}, \; {$^3$\tt tpsingh@tifr.res.in}

\bigskip

 {\it Published in  Mod. Phys. Lett. (2020) 2150019}
 
 {DOI:  https://doi.org/10.1142/S021773232150019X }

\end{center}

\bigskip

\centerline{\bf ABSTRACT}
\noindent  We have recently proposed a Lagrangian in trace dynamics, to describe a possible unification of gravity, Yang-Mills fields, and fermions, at the Planck scale. This Lagrangian for the  unified entity - called the aikyon - is invariant under global unitary transformations, and as a result possesses a novel conserved charge, known as the Adler-Millard charge. In the present paper, we derive an eigenvalue equation, analogous to the time-independent Schr\"{o}dinger equation, for the Hamiltonian of the theory. We show that in the emergent quantum theory, the energy eigenvalues of the aikyon are characterised in terms of a fundamental frequency times Planck's constant. The eigenvalues of this equation can, in principle, determine the values of the parameters of the standard model. We also report a ground state, in this theory of spontaneous quantum gravity, which could characterise a non-singular initial epoch in quantum cosmology. 

%The structure of our Lagrangian predicts a fourth generation of yet to be discovered eight new fermions.

\bigskip

\bigskip
\section{An eigenvalue equation for the hamiltonian of an aikyon}
We have recently proposed \cite{MPSingh, Singhspin} the following Lagrangian and action  in trace dynamics \cite{Adler:04, Adler:94, AdlerMillard:1996}, to describe the unification of gravity, Yang-Mills fields, and fermions:
\[
 \frac{S}{C_0} = \frac{1}{2}\int \frac{d\tau}{\tau_{Pl}}\; Tr \biggl[\biggr. \dfrac{L_{p}^{2}}{L^{4}} \biggl\{\biggr. i\alpha \biggl(\biggr. q_{B} + \dfrac{L_{p}^{2}}{L^{2}}\beta_{1} q_{F} \biggl.\biggr) + L \biggl(\biggr. \dot{q}_{B} + \dfrac{L_{p}^{2}}{L^{2}}\beta_{1}\dot{q}_{F} \biggl.\biggr) \biggl.\biggr\}\times \nonumber\\
\biggl\{\biggr. i\alpha \biggl(\biggr. q_{B} + \dfrac{L_{p}^{2}}{L^{2}}\beta_{2} q_{F} \biggl.\biggr) + L \biggl(\biggr. \dot{q}_{B} + \dfrac{L_{p}^{2}}{L^{2}}\beta_{2} \dot{q}_{F} \biggl.\biggr) \biggl.\biggr\} \biggl.\biggr]
\label{funlag}
\]
Here, the action is made dimensionless by scaling with respect to $C_0$, a real constant with dimensions of action, which is subsequently identified as being proportional to  Planck's constant (the proportionality constant being chosen so as to recover the correct numerical factor in the emergent Einstein's  equations).  $q_B$ and $q_F$ are respectively even-grade (hence bosonic)  and odd-grade (hence fermionic) Grassmann matrices, which describe Yang-Mills fields and their fermionic current source. Their time derivatives with respect to the Connes time $\tau$, i.e. $\dot{q}_B$  and $\dot{q}_F$, represent gravity and its fermionic mass source, respectively. Together, these four dynamical variables describe an `atom' of space-time matter, or an `aikyon', which has an associated length scale $L$. $\beta_1$ and $\beta_2$ are constant  fermionic matrices, and $\alpha$ is the gauge coupling constant. At the Planck scale, the universe is populated by many such aikyons, with the total action for the theory being the sum of their individual actions. At scales below Planck scale, quantum field theory and classical general relativity are emergent from the Planck scale trace dynamics of these aikyons. These aspects have been studied in detail in our recent papers \cite{MPSingh, Singhspin, Singh2020DA}. 
In particular, we have investigated the potential for the above Lagrangian to describe the unification of gravity with the standard model. The reader is referred to these two papers for details of these investigations.

In the present paper we derive an eigenvalue equation for the Hamiltonian of the aikyon, and suggest that this equation could in principle predict values of the standard model parameters. This equation turns out to be an analogue of the time-independent Schr\"{o}dinger equation in a potential, for four dynamical variables. In order to arrive at this equation, we first define two bosonic variables $q_1$ and $q_2$ as follows:
\begin{equation}
q_1 = q_B + a_0 \beta_1 q_F , \qquad q_2 = q_B + a_0 \beta_2 q_F
\end{equation}
where $a_0 \equiv L_P^2 / L^2$.
In terms of these two variables, the above trace Lagrangian for the aikyon can be written as
\begin{equation}
Tr {\cal L} = \frac{1}{2} a_1 a_0  Tr \left[\dot{q}_1  \dot{q}_2  - \frac{\alpha^2 c^2}{L^2} q_1 q_2 + i\frac{\alpha c}{L} \left(\dot{q}_1 q_2 + q_1 \dot{q}_2 \right) \right]
\label{lagnew}
\end{equation}
where $a_1 \equiv C_0 /cL_P $, which would be proportional to Planck mass once $C_0$ is identified to be proportional to Planck's constant $\hbar$. [The constant of proportionality  is chosen to ensure the correct numerical factors in the emergent Einstein equations. It is understood that the action principle is now being written as $S=\int d\tau\;{\cal L}$]. In writing this expression for the Lagrangian  a $c$ factor has been pulled out from the original definition of the `dot' time derivative. Henceforth, the time derivative of $q$ has dimensions of velocity, and since $a_1$ has dimensions of mass, the trace Lagrangian has the correct dimensions of energy. The last term in the trace Lagrangian is a total time derivative, and hence does not contribute to the equations of motion, and is henceforth dropped, so that we work with the Lagrangian:
\begin{equation}
Tr {\cal L} = \frac{1}{2} a_1 a_0  Tr \left[\dot{q}_1  \dot{q}_2  - \frac{\alpha^2 c^2}{L^2} q_1 q_2  \right]
\label{q1q2lag}
\end{equation}

The canonical momenta are given by
\begin{equation}
p_1 = \frac{\delta Tr {\cal  L}} {\delta \dot{q}_1}  = \frac{1}{2} a_1 a_0 \dot{q}_2 ; \qquad  p_2 = \frac{\delta Tr {\cal  L}} {\delta \dot{q}_2}  = \frac{1}{2} a_1 a_0 \dot{q}_1
\end{equation}
The Euler-Lagrange equations of motion are
\begin{equation}
\ddot{q}_1 = - \frac{\alpha^2 c^2}{L^2} q_1; \qquad \ddot{q}_2 = - \frac{\alpha^2 c^2}{L^2} q_2
\end{equation}
In terms of these two complex variables, the aikyon behaves like two independent complex-valued oscillators. However, the degrees of freedom of the aikyon couple with each other when expressed in terms of the self-adjoint variables $q_B$ and $q_F$. This is because $q_1$ and $q_2$ both depend on $q_B$ and $q_F$, the difference being that $q_1$ depends on $\beta_1$ and $q_2$ depends on $\beta_2$.

The trace Hamiltonian for an aikyon is 
\begin{equation}
Tr{\cal H} = Tr [p_1 \dot{q}_1 + p_2 \dot{q}_2 - Tr {\cal  L}] = \frac{a_1 a_0}{2}Tr \left[ \frac{4}{a^2_1 a^2_0}p_1 p_2 + \frac{\alpha^2 c^2}{L^2} q_1  q_2 \right]
\end{equation}
and Hamilton's equations of motion are
\begin{equation}
\dot{q}_1  = \frac{2}{a_1 a_0} p_2  \qquad  \dot {q}_2  = \frac{2}{ a_1 a_0 } p_1  ;  \qquad
\dot{p}_1 = -\frac{a_1 a_0 \alpha^2  c^2}{2L^2} q_2; \qquad \dot{p}_2 = -\frac{a_1 a_0 \alpha^2  c^2}{2L^2} q_1
\end{equation}
%This trace Hamitonian, though bosonic, is not real in general, and its real and imaginary parts are given by
%\begin{equation}
%\begin{split}
%Tr{\cal H^{SA}} =  Tr \left[ \frac{1}{a_1 a_0}p_1 p_2 - i\frac{\alpha c}{L} (p_1 q_1 + p_2 q_2) \right]
%\end{split}
%\end{equation}
Because the trace Lagrangian is invariant under global unitary transformations, it possesses a novel conserved charge, known as the Adler-Millard charge, and defined as follows:
\begin{equation}
\tilde{C} = \sum_i [q_{Bi}, p_{Bi}] - \sum_i\{q_{Fi,} p_{Fi}\}
\end{equation}
which is the sum, over all bosonic degrees of freedom, of the commutator $[q_B, p_B]$ minus the sum, over all fermionic degrees of freedom, of the 
anti-commutator $\{q_F, p_F\}$. This charge plays an important role in the theory, as we see below.

In our theory, the Dirac operator $D_B$ is defined as $D_B\equiv (1/L) dq_B/d\tau$, and this operator coincides with the standard Dirac operator on a Riemannian manifold when space-time emerges in the classical limit of the theory. The fermionic Dirac operator $D_F$ is defined as time derivative of $q_F$ so that together we have $D\equiv D_B + D_F$
where
\begin{equation}
D_B \equiv \frac{1}{Lc}\;  \frac{dq_B}{d\tau}; \qquad   D_F \equiv \frac{L_P^2}{L^2}\frac{\beta_1+\beta_2}{2Lc}\;  \frac{dq_F}{d\tau}
\end{equation}
When Yang-Mills fields and their fermionic charge source are included in the theory, we define the generalised Dirac operator $D_{newi}\equiv D_{Bnewi} + D_{Fnewi}$ where
\[
D_{Bnewi} = \dfrac{1}{Lc} \dot{\widetilde{Q}}_{B} \qquad and \qquad D_{Fnewi} = \frac{L_P^2}{L^2}\frac{\beta_1+\beta_2}{2Lc} \dot{\widetilde{Q}}_{F}
\label{eq:ddirac}
\]
and
\begin{equation}
{\dot{\widetilde{Q}}_B} =  (i\frac{\alpha c}{L} q_B +  \dot{q}_B); \qquad  {\dot{\widetilde{Q}}_F} =  (i\frac{\alpha c}{L} q_F +  \dot{q}_F);
\end{equation}
Having transformed to the variables $q_1$ and $q_2$ it is straightforward to show that $D_{newi}=D_1 + D_2$ where
\begin{equation}
D_1 = \frac{1}{2Lc} \bigg(\dot{q}_1 +i\frac {\alpha c}{L} q_1 \bigg)\ ; \qquad D_2 = \frac{1}{2Lc} \bigg(\dot{q}_2 +i\frac {\alpha c}{L} q_2 \bigg)
\end{equation}

In order to arrive at the desired Schr\"{o}dinger equation, if we divide $q_1$ and $q_2$ into their self-adjoint and anti-self-adjoint parts  as $q_1 = q_1^s + q_1^{as}$, $q_2 = q_2^s + q_2^{as}$, then we can use the self-adjoint and anti-self-adjoint parts as new dynamical variables:
\begin{equation}
q_1^s= \frac{1}{2}(q_1 + q_1^{\dagger}) ; \qquad q_1^{as} = \frac{1}{2}(q_1-q_1^{\dagger}) ; \qquad q_2^s= \frac{1}{2}(q_2 + q_2^{\dagger})  ;  \qquad q_2^{as}= \frac{1}{2}(q_2 - q_2^{\dagger})
\end{equation}
The Lagrangian then becomes
\begin{equation}
{\cal L} =  \frac{a_1 a_0}{2} Tr \bigg[ \frac{4}{a^2_1 a^2_0} \bigg\{ \dot{q}_2^s \dot{q}_1^s  + \dot{q}_2^{as} \dot{q}_1^{as} \bigg\}  
-\frac{\alpha^2 c^2}{L^2} \bigg\{q_1^s q_2^s + q_1^{as} q_2^{as} \bigg\} +   \frac{4}{a^2_1 a^2_0} \bigg\{ \dot{q}_2^s \dot{q}_1^{as}  + \dot{q}_2^{as} \dot{q}_1^{s} \bigg\}  
-\frac{\alpha^2 c^2}{L^2} \bigg\{q_1^s q_2^{as} + q_1^{as} q_2^{s}  \bigg\} \bigg]
\label{waahwaah}
\end{equation}
In terms of these new variables, and in terms of the self-adjoint and anti-self-adjoint parts of the momenta $p^s$ and $p^{as}$, 
the real and imaginary parts of the trace Hamiltonian are given as follows:
\begin{equation}
Tr H^s = \frac{a_1 a_0}{2} Tr \bigg[ \frac{4}{a^2_1 a^2_0} \bigg\{ p_1^s p_2^s  + p_1^{as} p_2^{as} \bigg\}  
+\frac{\alpha^2 c^2}{L^2} \bigg\{q_1^s q_2^s + q_1^{as} q_2^{as}  \bigg\} \bigg]
\label{h1}
\end{equation}
\begin{equation}
Tr H^{as} = \frac{a_1 a_0}{2} Tr \bigg[ \frac{4}{a^2_1 a^2_0} \bigg\{ p_1^s p_2^{as}  + p_1^{as} p_2^{s} \bigg\}  
+\frac{\alpha^2 c^2}{L^2} \bigg\{q_1^s q_2^{as} + q_1^{as} q_2^{s}  \bigg\} \bigg]
\label{him}
\end{equation}

The conserved Adler-Millard charge, and its self-adjoint and anti-self-adjoint components are
\begin{equation}
\begin{split}
&\tilde{C}=[q_1, p_1] + [q_2, p_2]\\
&\tilde{C}^{as} = [q_1^s, p_1^{s}] + [q_1^{as}, p_{1}^{as}] +  [q_2^s, p_2^{s}] +  [q_2^{as}, p_2^{as}] \\
& \tilde{C}^{s} = [q_1^s, p_1^{as}] + [q_1^{as}, p_{1}^{s}] +  [q_2^s, p_2^{as}] +  [q_2^{as}, p_2^{s}]
\end{split}
\end{equation}
In the theory of trace dynamics, at energies below Planck scale, quantum theory (without a background spacetime) emerges. This is achieved by coarse-graining the theory over many Planck time scales, and by applying the methods of statistical thermodynamics to arrive at the emergent quantum theory. This happens provided the self-adjoint part of the Adler-Millard charge can be neglected, and the anti-self-adjoint part of the Hamiltonian can be neglected. When that happens, the Adler-Millard charge gets equipartitioned over the four degrees of freedom, the equipartitioned value is identified with Planck's constant, and quantum commutation relations emerge, for the statistically averaged dynamical variables at statistical equilibrium:
\begin{equation}
[q_1^s, p_1^{s}]=i\hbar, \qquad [q_1^{as}, p_{1}^{as}]=i\hbar, \qquad [q_2^s, p_2^{s}] = i\hbar, \qquad [q_2^{as}, p_2^{as}]=i\hbar
\end{equation}
The averaged dynamical variables obey Heisenberg equations of motion. The above commutation relations allow us to write the momenta as displacement operators conjugate to their corresponding configuration variables, enabling us to move to the position representation
\begin{equation}
p_1^{s} = -i\hbar \frac{\partial \ }{\partial q_1^s},\qquad p_1^{as} = -i\hbar \frac{\partial \ }{\partial q_1^{as}},\qquad p_2^{s} = -i\hbar \frac{\partial \ }{\partial q_2^s},\qquad   p_2^{as} = -i\hbar \frac{\partial \ }{\partial q_2^{as}},
\end{equation}
Using these, and the form of the self-adjoint part of the Hamiltonian operator, we arrive at the following sought after eigenvalue equation $H^s\psi = \lambda\psi$ for the operator Hamiltonian for these four  dynamical variables [with the knowledge that in the emergent quantum theory, the configuration variables are treated as c-numbers, in the position representation]:
\begin{equation}
- \frac{2\hbar^2}{a_1 a_0} \left( \frac{\partial^2\ }{\partial q_1^s \; \partial q_2^s}  +\frac{\partial^{2}\ }{\partial q_1^{as} \; \partial q_2^{as}} \right) \psi  + \frac{\alpha^2 c^2 a_1 a_0 }{2L^2} (q_1^s q_2^s + q_1^{as}q_2^{as} )\psi  = \lambda \psi
\label{schofun}
\end{equation} 
This fundamental equation is a time-independent Schr\"{o}dinger equation for these four variables, in the potential shown in the equation. It is the analog of the Wheeler-deWitt equation in our theory, and predicts the quantum gravitational states of the aikyon. The ground state solution of this equation will tell us about the non-singular state of the universe at the big bang epoch. Hopefully, this equation also carries information about the parameters of the standard model, considering that it depends on the gauge coupling constant $\alpha$, and also on the length parameter $L$ from which the fermionic mass $m=\hbar / Lc$ is defined. These investigations are currently in progress. 

Here, we include a few preliminary observations about this equation and its solutions. The system has the characteristics of an oscillator, considering the bi-linear nature of the potential. That suggests we could think of the coefficient of the potential term as one-half of a `spring constant' k, hence $k = \alpha^2 c^2 a_1 a_0/L^2$. Also, from the kinetic energy term, the `mass' $m$ can be read off as $m = a_1 a_0/4$  which is proportional to $ m_{Pl}L_P^2 /L^2$. This would imply that the `frequency' $\omega = \sqrt{k/m}$ is given by $\omega = 2\alpha c/L$. The most remarkable thing here is that the Planck scale has dropped out from the estimate of the frequency! We comment below on the possible implications of this result. A scale analogous to this frequency is actually seen in the solution of the above Schr\"{o}dinger equation, though with a twist.

To see this, we start by noting that the above equation for the two self-adjoint and  two anti-self adjoint variables can be separated into two equations, by writing the wave-function $\psi(q_1^s, q_2^s, q_1^{as}, q_2^{as} )$ as a product state:
\begin{equation}
\psi(q_1^s, q_2^s, q_1^{as}, q_2^{as} ) \equiv \psi_{ss} (q_1^s, q_2^s) \;\psi_{aa}(q_1^{as} q_2^{as})
\end{equation}
Substituting this in the above Schr\"odinger equation yields two decoupled Schr\"{o}dinger equations:
\begin{equation}
- \frac{2\hbar^2}{a_1 a_0} \left( \frac{\partial^2\ }{\partial q_1^s \; \partial q_2^s} \right) \psi_{ss}  + \frac{\alpha^2 c^2 a_1 a_0 }{2L^2} (q_1^s q_2^s  )\psi _{ss} = \lambda_{ss} \psi_{ss}
\label{scho1}
\end{equation} 
\begin{equation}
- \frac{2\hbar^2}{a_1 a_0} \left( \frac{\partial^2\ }{\partial q_1^{as} \; \partial q_2^{as}} \right) \psi_{aa}  + \frac{\alpha^2 c^2 a_1 a_0 }{2L^2} (q_1^{as}q_2^{as} )\psi_{aa}  = \lambda_{aa} \psi_{aa}
\end{equation} 
with $\lambda = \lambda_{aa} + \lambda_{ss}$. It turns out that the solutions are not oscillatory, but exponentially decaying [expected, for the `spatial part' of a system bounded by a potential, the oscillatory time-dependence in Connes time is in any case there for this stationary solution] or exponentially growing. We will find the eigenvalue spectrum and the corresponding eigenstates in the next section. 

The imaginary part of the trace Hamiltonian is given by Eqn. (\ref{him}) above. It has contributions from cross-terms which couple the self-adjoint parts of position and momenta to their anti-self-adjoint parts. This anti-self-adjoint part of the Hamiltonian is expected to become significant when sufficiently many aikyons get entangled with each other. When that happens, superpositions of position states become extremely short-lived and spontaneous localisation results, in the spirit of the Ghirardi-Rimini-Weber-Pearle theory of objective collapse \cite{Ghirardi:86, Pearle:89,Ghirardi2:90, Bassi:03, RMP:2012}. Hence we expect to be able to derive collapse models, including the collapse rate parameter, and the spectrum of the collapse noise, starting from the Lagrangian of our theory. This work is currently in progress. One can expect the collapse rate parameter to be related to $\alpha c/L$, the frequency that arises naturally in our theory.

Next, we present the exact spectrum of the eigenvalues of the above Schr\"{o}dinger equation, which confirms the above preliminary observations.

\subsection{Spectrum of eigenvalues for the energy eigenstates of the aikyon}
The Schr\"odinger equation (\ref{scho1}) is a partial differential equation in two variables, of the form
\begin{equation}
\frac{\partial^2\psi}{\partial x \partial y} + K_1 xy\psi = K_2\psi
\label{fundif}
\end{equation}
where
\begin{equation}
K_1 =  - \frac{a_1^2 a_0^2 \alpha^2 c^2}{4L^2\hbar^2}; \qquad K_2 = - \frac{\lambda a_1 a_0}{ 2\hbar^2}
\end{equation}
where $\lambda$ is the eigenvalue of the Schr\"{o}dinger equation, and we will assume $xy\geq0$ so that the potential is bounded from below.

Let us consider the following ansatz for the solution to this equation:
\begin{equation}
\psi(x,y) \sim f(x) g(y) \exp [- mxy]
\end{equation}
where $f(x)$ is an arbitrary function of $x$ and $g(y)$ is an arbitrary function of $y$, and $m>0$ so that the wave-function is bounded.

By differentiating the wave function in this ansatz first with respect to $y$ and then with respect to $x$, and substituting this second order derivative in (\ref{fundif}) we get the following equation:
\begin{equation}
\begin{split}
f'(x) g'(y) - & myf(x) g'(y) -  mx f'(x) g(y) - m f(x) g(y) + m^2 xy f(x) g(y) \\
&+ K_1 xy f(x) g(y)=K_2 f(x) g(y)
\end{split}
\end{equation} 
Now comparing both sides we get two conditions:
\begin{equation}
 \qquad m^2 + K_1 =0 \implies m= \pm (-K_1)^{1/2} \implies m = \frac{a_1 a_0 \alpha c}{2L\hbar}
\label{con2}
\end{equation}
where we have ignored the negative root because we have assumed $m>0$.
Second condition is
\begin{equation}
 \quad \frac{f'(x)g'(y)}{f(x)g(y)}- m\bigg[\frac{yg'(y)}{g(y)} +  \frac{x f'(x)}{f(x)} +  1\bigg] 
=K_2 
\label{con3}
\end{equation}
Since $K_2$ determines the eigenvalues, LHS of this equation must be independent of $x$ and $y$. For this to be satisfied we must have the following;
\begin{equation}
f'(x)g'(y) = 0
\end{equation}
which implies that both $f'(x)$ and $g'(y)$ cannot be non zero simultaneously.We basically have two cases for Eqn.(\ref{con3}).
\begin{equation}
{\rm Case\ I}:\qquad g'(y) = 0 \implies g(y) = {\rm cons.}\qquad {\rm and}    \qquad  \frac{x f'(x)}{f(x)}=r\implies f(x) = x^r
\end{equation}
where we require $r\geq0$ for wavefunction to be physical. 
\begin{equation}
{\rm Case\ II}:\qquad f'(x) = 0 \implies f(x) = {\rm cons.}\qquad {\rm and}    \qquad  \frac{y g'(y)}{g(y)}=s\implies g(y) = y^s
\end{equation}
where $s\geq0$. Therefore the the general form of eigenstates for Eqn. (\ref{fundif}) will be
\begin{equation}
\psi_{rs} = Cx^r y^s \exp \bigg[ - \frac{a_1 a_0 \alpha c}{ 2L\hbar}xy \bigg]
\end{equation}
with eigenvalue
\begin{equation}
\lambda = \frac{\hbar\alpha c}{L} \; (1+r+s)
\end{equation}
where one of $r$ and $s$ must be zero.
The ground state is given by
\begin{equation}
\psi_{00} = C_0 \exp \bigg[ - \frac{a_1 a_0 \alpha c}{ 2L\hbar}xy \bigg]; \qquad \lambda = \frac{\hbar \alpha c}{L}
\end{equation}
It has the character of a non-singular ground state, with a `minimum area spread' given by $L\hbar / a_1 a_0 \alpha c = L^3/6\pi \alpha L_P$. [We have taken $a_1=6\pi m_{Pl}$, this coefficient being the correct one for recovering Einstein equations with correct numerical factors].

We do not yet have  reason to assume that $r$ and $s$ take only discrete values. Therefore we have a continuous spectrum of energy eigenstates. Since one of $r$ and $s$ must be zero, for each  $r\geq0$ we have doubly degenerate excited states each having an eigenvalue $(1+r)\hbar \alpha c/L$:
\begin{equation}
\psi_{r0} = C_r \; x^r \exp \bigg[ - \frac{a_1 a_0 \alpha c}{ 2L\hbar}xy \bigg]; \qquad     \psi_{0r}= C_r\; y^r\exp \bigg[ - \frac{a_1 a_0 \alpha c}{ 2L\hbar}xy \bigg]
\end{equation}

For the original Schr\"odinger Equation (\ref{schofun}) one can now easily deduce, assuming that $q_1^{as}q_2^{as}$ is also positive, the ground state is given by
\begin{equation}
\psi_0 = C_0  \exp \bigg[ - \frac{a_1 a_0 \alpha c}{ 2L\hbar} (q_1^sq_2^s+q_1^{as}q_2^{as}) \bigg]
\end{equation}
and has the energy eigenvalue $2\hbar \alpha c/L$. This is the ground state of an aikyon in our theory. If the aikyon is an electron, this state non-perturbatively describes the electron, together with its mass aspect, and with all the fields that it produces! We believe this is the first ever result of its kind. Again there is a minimum area spread given by $L^3/6\pi \alpha L_P$, and suggests a non-singular initial state of the universe at the epoch of its beginning.

The excited states of the aikyon can be written as
\begin{equation}
\psi = C \psi_{r1, s1} (q_1^s, q_2^s) \; \psi_{r2, s2}(q_1^{as} q_2^{as})
\end{equation}
and have the eigenvalue 
\begin{equation}
\lambda = \frac{\hbar \alpha c}{L} (2 + r_1 + s_1 + r_2 + s_2)
\end{equation}
where one of $r_1$ or $s_1$ must be zero, and one of $r_2$ or $s_2$ must be zero.

It remains to be investigated if these solutions are relevant in the quantum cosmological context, and if they can describe an expanding universe at the big bang epoch and address the issue of singularity avoidance. What is highly significant is that the energy eigenvalue is independent of Planck scale. We recall that in our theory the underlying trace dynamics is assumed to take place at the Planck scale. Whereas the quantum gravity theory whose solutions we are describing here emerges after coarse-graining over many Planck time scales. However, in our theory there is nothing special about Planck energy scale. We could have chosen any other sufficiently high energy scale at which quantum field theory has not been tested, and at which scale we assume the laws of quantum field theory to be replaced by those of trace dynamics. This suggests that the ratio $\alpha/L$ which appears in the energy eigenvalue remains unchanged during coarse-graining. Since coarse-graining is a mechanism for coming to lower energy scales from higher energy scales, and since the gauge coupling constant $\alpha$ runs with energy, a change in $\alpha$ would imply a change in $L$. And hence a change in bare mass, since $L$ is identified with Compton wavelength; thus it might be possible to estimate lepton and quark bare masses as well as standard model parameters, from a more careful and detailed analysis of this theory. This will be attempted during forthcoming investigations, and we hope that other researchers will also find it interesting to explore this avenue.

\section{An aikyon in octonionic space}
It is possible that the  energy eigenvalues that we have found above for the aikyon, relate to the standard model in a novel way.
In our recent work \cite{Singhspin} we have proposed that the aikyon lives and evolves in an 8-D octonionic space. If this were to be so, the fundamental Lagrangian (\ref{funlag}), reproduced below,
\[
 \frac{S}{C_0} = \frac{1}{2}\int \frac{d\tau}{\tau_{Pl}}\; Tr \biggl[\biggr. \dfrac{L_{p}^{2}}{L^{4}} \biggl\{\biggr. i\alpha \biggl(\biggr. q_{B} + \dfrac{L_{p}^{2}}{L^{2}}\beta_{1} q_{F} \biggl.\biggr) + L \biggl(\biggr. \dot{q}_{B} + \dfrac{L_{p}^{2}}{L^{2}}\beta_{1}\dot{q}_{F} \biggl.\biggr) \biggl.\biggr\}\times \nonumber\\
\biggl\{\biggr. i\alpha \biggl(\biggr. q_{B} + \dfrac{L_{p}^{2}}{L^{2}}\beta_{2} q_{F} \biggl.\biggr) + L \biggl(\biggr. \dot{q}_{B} + \dfrac{L_{p}^{2}}{L^{2}}\beta_{2} \dot{q}_{F} \biggl.\biggr) \biggl.\biggr\} \biggl.\biggr]
\label{funlag2}
\]
could describe the standard model of particle physics, and its  unification with gravity. This trace Lagrangian is invariant under global unitary transformations, and in the octonionic space these are generated by the algebra of automorphisms which form the group $G_2$. The symmetries are described by the Clifford algebra $Cl(2)$ generated from the algebra of complex quaternions, and the Clifford algebra $Cl(6)$ generated from the algebra of complex octonions \cite{f1, f3}.
In Lagrangian (\ref{funlag2}), the objects $q_B$ and $\dot{q}_B$ together form an octonion, and similarly $q_F$ and $\dot{q}_F$ form another octonion. Every one of $q_B, q_F, \dot{q}_B, \dot{q}_F$ carries four indices, with the time derivative terms carrying the octonionic indices $(1, e_1, e_2, e_3)$ and $q_B$ and $q_F$ each carrying the other four octonionic indices  $(e_4, e_5, e_6, e_7)$. Hence we have 
\begin{equation}
\begin{split}
q_B=(q_{Be4}, q_{Be5}, q_{Be6}, q_{Be7}); \qquad \dot{q}_B = (\dot{q}_{B1}, \dot{q}_{Be1}, \dot{q}_{Be2}, \dot{q}_{Be3})\\
q_F=(q_{Fe4}, q_{Fe5}, q_{Fe6}, q_{Fe7}); \qquad \dot{q}_F = (\dot{q}_{F1}, \dot{q}_{Fe1}, \dot{q}_{Fe2}, \dot{q}_{Fe3})
\end{split}
\end{equation}
Thus, when the brackets are opened up, the Lagrangian has 256 terms (some of these will be removed as they are total time derivatives) and these could incorporate the standard model and gravity \cite{Singhspin}. $q_B$ and $q_F$ respectively describe would-be-Yang-Mills and its fermionic charged source. $\dot{q}_B$ and $\dot{q}_F$ describe would-be-gravity and its fermionic mass source. Together these four entities describe the unification of gravity and Yang-Mills, with each other and with the fermionic source, so as to form the aikyon, whose Lagrangian  is the one written above.

The structure of the Lagrangian (\ref{funlag2}) in our theory then suggests it can describe one generation of fermions.This can be inferred by noting that the space-time symmetries of the charge aspect of the fermions arise from the cross-terms $i\alpha  \beta_1 q_F \times L\dot{q}_B$ and 
$L\dot{q}_B \times i\alpha \beta_2 q_F$. In the octonionic space spanned by the directions $(1, e_1, e_2, e_3, e_4, e_5, e_6, e_7)$ the would-be space-time directions are $(1, e_1, e_2, e_3)$ and the other four directions are internal directions. The Dirac operator $D_B \propto \dot{q}_B$ has four space-time components and these are $[D_{B1}, D_{Be1}, D_{Be2}, D_{Be3}]$. The fermionic charge aspect has the four components $[q_{Fe4}, q_{Fe5}, q_{Fe6}, q_{Fe7}]$. Therefore, the cross terms, between them, have $4\times 4 + 4\times 4 =32$ components. Suppressing the labels $D_F$ and $q_F$ these thirty-two components can be represented as
\begin{equation}
\beta_1 \; [e_4, e_5, e_6, e_7]\times [1, e_1, e_2, e_3] \quad + \quad [1, e_1, e_2, e_3]\times \beta_2 \; [e_4, e_5, e_6, e_7]
\end{equation}
Every one of the four directions $(e_4, e_5, e_6, e_7)$ gives rise to  eight terms, as can be easily read off from this previous expression. Thus these thirty-two components can also be equivalently written as four sets of eight terms each, as follows:
\begin{equation}
\begin{split}
[\beta_1 e_4 \times 1, \beta_1 e_4 \times e_1, \beta_1 e_4 \times e_2, \beta_1 e_4 \times e_3]  \quad & + \quad [1\times \beta_2 e_4, e_1\times\beta_2 e_4, e_2 \times\beta_2 e_4, e_3 \times \beta_2 e_4] \\
[\beta_1 e_5 \times 1, \beta_1 e_5 \times e_1, \beta_1 e_5 \times e_2, \beta_1 e_5 \times e_3]  \quad & + \quad [1\times \beta_2 e_5, e_1\times\beta_2 e_5, e_2 \times\beta_2 e_5, e_3 \times \beta_2 e_5] \\
[\beta_1 e_6 \times 1, \beta_1 e_6 \times e_1, \beta_1 e_6 \times e_2, \beta_1 e_6 \times e_3]  \quad & + \quad [1\times \beta_2 e_6, e_1\times\beta_2 e_6, e_2 \times\beta_2 e_6, e_3 \times \beta_2 e_6] \\
[\beta_1 e_7 \times 1, \beta_1 e_7 \times e_1, \beta_1 e_7 \times e_2, \beta_1 e_7 \times e_3]  \quad & + \quad [1\times \beta_2 e_7, e_1\times\beta_2 e_7, e_2 \times\beta_2 e_7, e_3 \times \beta_2 e_7] 
\end{split}
\end{equation}
The eight terms in any one row represent the coupling of one internal direction to the four would-be-spacetime directions.
It appears that there are two leptons (as desired), these being the  components 
of the type $(\beta_1 e_4 \times 1, 1\times \beta_2 e_4)$, and the other twenty-four components possibly represent the six quarks. It could be that a more direct interpretation arises in terms of the $q_1$ and $q_2$ variables studied in the previous section.

The vector bosons come from the square term $-\alpha^2 q_B^2$ in the Lagrangian, which lives in the internal space. So $q_B$ carries the four octonionic indices 
$(e_4, e_5, e_6, e_7)$. Hence there are sixteen components to this square term,  $-\alpha^2 q_B^2$. We expect that the twelve off-diagonal components represent the twelve massless vector bosons, whereas the four diagonal components presumably represent the four Higgs bosons (prior to spontaneous symmetry breaking).
Hence there are a total of sixteen bosons eight  fermions of one generation the theory.

Where do the heavier fermion generations come from? It is possible that there arises in the theory an eigenvalue equation for the fermion masses, whose unequal roots give rise to different fermion masses. Possibly, this information is already contained in the analysis of the previous section. Further developments in this direction are reported in \cite{Singh2020DA}. 

%Judging by the structure of the theory, we would like to conjecture that there is a fourth yet undiscovered generation of eight new fermions. It is almost impossible to see how the number three [which is the number of known generations] can arise in the scenario considered in this paper.

\subsection{Understanding the Lagrangian in octonionic space}
We will assume for the sake of further investigation that the Grassmann matrix $q$ is defined over the field of octonions. The space in which it resides is 8-D, and labelled conventionally by the directions $(1, e_1, e_2, e_3, e_4, e_5, e_6, e_7)$. The first direction is the real line and the other seven are imaginary directions whose algebra is described by the multiplication rules of the Fano plane. The following index allocation is assumed for the variables $q_B, \dot{q}_B, q_F$ and $\dot{q}_F$. $\dot{q}_B$ has quaternionic indices $(1, e_1, e_2, e_3)$ where $(e_1, e_2, e_3)$ is a quaternionic triplet, so that these four directions form a closed algebra. Therefore $\dot{q}_B$ is a quaternion, and the term $\dot{q}_B^2$ in the trace Lagrangian describes would-be emergent spacetime. This term has 16 components. The term $iq_B$ carries the octonionic indices $(e_4, e_5, e_6, e_7)$. Its square has 16 components and it describes the standard model gauge fields and the four Higgs bosons. One can think of the four spacetime directions [could be thought of as `horizontal' \cite{Singhspin}] as being orthogonal to the four internal gauge directions [could be thought of as `vertical']. 

The Grassmann matrix $\dot{q}_F$ is assumed to have the quaternionic indices $(1, e_1, e_2, e_3)$ and describes the mass aspect of the fermionic degrees of freedom. It's square term is order $L_P^6$ and does not contribute in the classical limit, which comes from terminating the heat kernel expansion at order $L_P^4$. This term provides quantum corrections to the classical theory, and we need not consider these in the classical limit. However, for reasons to be mentioned below, we retain them for now.  Further, in the trace Lagrangian,  cross terms involving  $\dot{q}_F$ contribute, but only cross terms with $\dot{q}_B$ contribute: there are 32 such terms. Other cross-terms involving $\dot{q}_F$ contribute to the total time derivative in the Lagrangian, and are hence dropped [more on this below]. We emphasize that $\dot{q}_B$ and $\dot{q}_F$ carry identical quaternionic indices: together  they form the `horizontal' gravity-mass quaternionic `plane' in octonioniic space; it is the domain of emergent 4-D space-time.

The Grassmann matrix $iq_F$ is assumed to have the the octonionic indices $(e_4, e_5, e_6, e_7)$; this represents fermionic charge aspect and together with $q_B$, which has the same indices, it forms the `vertical' gauge-charge `plane' in the octonionic space. Note that these four directions do not involve the real line (which is what makes them different from spacetime) and they do not form a closed algebra. Thus one can have space-time-mass without gauge-charge, but one cannot have gauge-charge without space-time. The contribution of $q_F$ to the Lagrangian, from its square, is order $L_P^6$ and is retained. Its contribution from product with $\dot{q}_B$ and $\dot{q}_F$ is  dropped, being part of total time-derivative. Thus it only contributes from terms of the form $q_F q_B$ and $q_B q_F$, there being a total of 32 such terms.

One can easily check from the Lagrangian (\ref{funlag}) and (\ref{lagnew}) that a total of 128 terms, out of the original 256, contribute to the total time derivatives, and are hence removed. There are  32 terms of order $L_P^6$ which are kept. That gives 128 terms which are as follows: 16 terms in $\dot{q}_B^2$ which describe gravity, 16 terms in $q_B^2$ which describe gauge-fields, 32 terms which describe gravity-mass interaction,  32 terms which describe gauge-charge interaction, and 32 fermion-squared terms of order $L_P^6$ whose curious status we comment on below.

The reduced trace Lagrangian with 128 terms can hence be written as
\begin{equation}
\begin{split}
 \frac{S}{C_0} = \frac{1}{2}\int \frac{d\tau}{\tau_{Pl}}\; Tr \bigg[\biggr.\frac{L_P^2}{L^2}\bigg\{\biggr.  \dot{q}_B^2 +\frac{L_P^2}{L^2} \dot{q}_B\beta_2 \dot{q}_F + \frac{L_P^2}{L^2} \beta_1 \dot{q}_F \dot{q}_B   + \frac{L_P^4}{L^4} \beta_1\dot{q}_F \beta_2\dot{q}_F  \\- \frac{\alpha^2}{L^2} \bigg(q_B^2     
+ \frac{ L_P^2}{L^2} q_B \beta_2 q_F + \frac{ L_P^2}{L^2} \beta_1 q_F q_B +  \frac{L_P^4}{L^4} \beta_1 q_F \beta_2 q_F  \bigg)
 \biggl. \bigg\} \biggl.\bigg] 
 \end{split}
 \label{fulllag}
 \end{equation}
Each of the eight terms has sixteen components, giving a total of 128, same as in the Lagrangian in Eqn. (\ref{q1q2lag}), there being 64 terms of kinetic energy type, and 64 terms of potential energy type. Although the $i$ factors have disappeared, there are anti-self-adjoint terms in the Lagrangian, as expected. The harmony between gravity and gauge-fields is noteworthy!

If we set $\alpha=0$, half of the terms vanish, and we are in the `horizontal' gravity-mass plane, which is quaternionic and has a closed algebra. This emerges as spacetime-gravity-mass in the classical limit [will have gauge-charge contributions if $\alpha$ is not zero]. The Lagrangian is
\begin{equation}
 \frac{S}{C_0} = \frac{1}{2}\int \frac{d\tau}{\tau_{Pl}}\; Tr \bigg[\frac{L_P^2}{L^2}\bigg\{ \dot{q}_B^2   
 +\frac{L_P^2}{L^2} \dot{q}_B\beta_2 \dot{q}_F + \frac{L_P^2}{L^2} \beta_1 \dot{q}_F \dot{q}_B  + \frac{L_P^4}{L^4} \beta_1\dot{q}_F \beta_2\dot{q}_F 
  \bigg\} \bigg] 
 \end{equation}
 This is the pure gravity case that we studied in \cite{maithresh2019}. If the matter part is ignored and if we consider only the first term $\dot{q}_B^2$, it is the complex quaternion part studied by Furey \cite{f1} to describe the Lorentz symmetry $Cl(2)$. When matter is included, it appears to suggest the Clifford algebra $Cl(4)$. When the full Lagrangian (\ref{fulllag}) is considered, the associated symmetry is described by $Cl(6)$ and this is the Lagrangian for one generation of fermions in the standard model, unified with gravity.
 
 We now  illustrate possible adjointness properties of this Lagrangian, after making some assumptions on the $\beta$ matrices. This choice is not unique but only illustrative. We assume $q_B$  and $q_F$ to be self-adjoint, and $\beta_1 q_F$ and $\beta_2 q_F$ to be each anti-self-adjoint.  With these conditions  we have
\begin{equation}
\begin{split}
&q_1^s = q_B, \ q_1^{as} = \beta_1\ q_F, \  q_2^s = q_B, \ q_2^{as} = \beta_2 q_F,  \\
& p_1^s = \frac{1}{2} a_1a_0 \dot{q}_B, \  p_2^s = \frac{1}{2} a_1a_0 \dot{q}_B, \
p_1^{as} =  \frac{1}{2} a_1a_0 \beta_2  \dot{q}_F, \  p_2^{as} =  \frac{1}{2} a_1a_0 \beta_1  \dot{q}_F
\end{split}
\end{equation}
In terms of these variables, the Lagrangian in (\ref{fulllag}) becomes
\begin{equation}
{\cal L} =  \frac{a_1 a_0}{2} Tr \bigg[ \frac{4}{a^2_1 a^2_0} \bigg\{ \dot{q}_2^s \dot{q}_1^s  + \dot{q}_2^{as} \dot{q}_1^{as} \bigg\}  
-\frac{\alpha^2 c^2}{L^2} \bigg\{q_1^s q_2^s + q_1^{as} q_2^{as} \bigg\} +   \frac{4}{a^2_1 a^2_0} \bigg\{ \dot{q}_2^s \dot{q}_1^{as}  + \dot{q}_2^{as} \dot{q}_1^{s} \bigg\}  
-\frac{\alpha^2 c^2}{L^2} \bigg\{q_1^s q_2^{as} + q_1^{as} q_2^{s}  \bigg\} \bigg]
\label{waah}
\end{equation}
which of course matches with (\ref{waahwaah}) obtained from the Lagrangian (\ref{q1q2lag}), as expected. The first half is the self-adjoint part and the second half is the anti-self-adjoint part. Each of the eight terms in the Lagrangian has sixteen components, which is true also for the Hamiltonian in Eqns. (\ref{h1}) and (\ref{him}). If we were to drop the 32 terms of order $L_P^6$ terms from the Lagrangian,  the self-adjoint  terms $p_1^{as} p_2^{as}$ and $q_{1}^{as} q_2^{as}$ would disappear from the above Lagrangian, and also alter the eigen-solutions we have found for the Hamiltonian operator. The curious role of these terms, which do not survive in the leading order classical limit, remains to be understood.

\section{Concluding Remarks}

Since the Hamiltonian has an anti-self-adjoint component, it implies that under  the circumstances when this component becomes significant, unitary evolution is lost, and spontaneous localisation takes place. Spontaneous localisation then acts as a means of suppressing four of the octonionic dimensions in the classical limit:
As we have noted, the aikyon lives in an 8-D octonionic space. Four of these constitute would-be-classical spacetime. Which four? This is very interesting indeed. The octonionic space has one real direction and seven imaginary directions. The rules of multiplication amongst the vectors along the imaginary direction are given by the Fano plane. The seven imaginary vectors criss-cross in triplets of three vectors such that any three in the same triplet form a closed algebra, often called quaternionic triplets. Pick a triplet - along with the real direction: these four form would-be-spacetime! It is a closed algebra. Would-be-gravity lives only in these four directions. The Yang-Mills fields are associated with the other four directions in the octonionic space; though more strictly they lie all over the octonionic space,  because the other four directions do not form a closed algebra. There is leakage from these directions onto the would-be-spacetime directions.

Aikyons evolve in this 8-D octonionic space. Perhaps we can talk of `collisions' in this space as a means of interaction. We can surely talk of entanglement of aikyon states. When sufficiently many of them get entangled, spontaneous localisation results [imaginary part of the Hamiltonian becomes significant], and entanglement breaks down. The entangled system localises to one of the states in the original superposition. Now, the real part of the Hamiltonian (to which the localisation takes place) depends {\it only} on the four directions which involve the real line and a quaternionic triplet [the would-be-spacetime]. Spontaneous localisation hence results in the emergence of a classical spacetime, with classical matter fields and classical gauge fields and gravity \cite{maithresh2019, stcw}.

In essence, spontaneous localisation suppresses dimensions. The classical system lives in 4-D spacetime. Whereas a quantum system lives in 8-D octonionic space. So there is no need to compactify the extra dimensions - classical systems never go there. Whereas quantum systems do, and that is perfectly fine. That is what makes them different from classical systems.

%\smallskip

\noindent {\bf Acknowledgements}: It is our pleasure to thank Sourabh Magare, Abhishek Sharma and Varun Srivastava for discussions and valuable help  on the eigenvalue problem for the Schr\"{o}dinger equation discussed in this paper.

%\section{Introduction}

\vskip 0.2 in
\centerline{\bf REFERENCES}
\bibliographystyle{unsrt}
\bibliography{biblioqmtstorsion}

\def\polhk#1{\setbox0=\hbox{#1}{\ooalign{\hidewidth
  \lower1.5ex\hbox{`}\hidewidth\crcr\unhbox0}}} \def\cprime{$'$}
  \def\cprime{$'$}
\begin{thebibliography}{10}

\bibitem{MPSingh}
Meghraj~M S, Abhishek Pandey, and Tejinder~P. Singh.
\newblock Why does the {Kerr-Newman} black hole have the same gyromagnetic
  ratio as the electron?
\newblock {\em submitted for publication}, arXiv:2006.05392, 2020.

\bibitem{Singhspin}
Tejinder~P. Singh.
\newblock Octonions, trace dynamics and non-commutative geometry: a case for
  unification in spontaneous quantum gravity.
\newblock {\em Zeitschrift f\"ur Naturforschung A}, DOI:
  https://doi.org/10.1515/zna-2020-0196:arXiv:2006.16274v2, 2020.

\bibitem{Adler:04}
Stephen~L. Adler.
\newblock {\em Quantum theory as an emergent phenomenon}.
\newblock Cambridge University Press, Cambridge, 2004.

\bibitem{Adler:94}
Stephen~L. Adler.
\newblock Generalized quantum dynamics.
\newblock {\em Nucl. Phys. B}, 415:195, 1994.

\bibitem{AdlerMillard:1996}
Stephen~L. Adler and Andrew~C. Millard.
\newblock Generalised quantum dynamics as pre-quantum mechanics.
\newblock {\em Nucl. Phys. B}, 473:199, 1996.

\bibitem{Singh2020DA}
Tejinder~P. Singh.
\newblock Trace dynamics and division algebras: towards quantum gravity and
  unification.
\newblock {\em Zeitschrift f\"ur Naturforschung A}, arXiv:2009.05574v44
  [hep-th]:DOI: https://doi.org/10.1515/zna--2020--0255, 2020.

\bibitem{Ghirardi:86}
Gian~Carlo Ghirardi, Alberto Rimini, and Tullio Weber.
\newblock Unified dynamics for microscopic and macroscopic systems.
\newblock {\em Phys. Rev. D}, 34:470--491, 1986.

\bibitem{Pearle:89}
Philip Pearle.
\newblock Combining stochastic dynamical state-vector reduction with
  spontaneous localization.
\newblock {\em Phys. Rev. A}, 39:2277, 1989.

\bibitem{Ghirardi2:90}
Gian~Carlo Ghirardi, Philip Pearle, and Alberto Rimini.
\newblock Markov processes in {Hilbert} space and continuous spontaneous
  localization of systems of identical particles.
\newblock {\em Phys. Rev. A}, 42:78--89, 1990.

\bibitem{Bassi:03}
Angelo Bassi and Gian~Carlo Ghirardi.
\newblock Dynamical reduction models.
\newblock {\em Phys. Rep.}, 379:257--426, 2003.

\bibitem{RMP:2012}
Angelo Bassi, Kinjalk Lochan, Seema Satin, Tejinder~P. Singh, and Hendrik
  Ulbricht.
\newblock Models of wave function collapse, underlying theories, and
  experimental tests.
\newblock {\em Rev. Mod. Phys.}, 85:471 arXiv:1204.4325 [quant--ph], 2013.

\bibitem{f1}
Cohl Furey.
\newblock Standard model physics from an algebra? {Ph. D.} thesis, university
  of {Waterloo}.
\newblock arXiv:1611.09182 [hep-th], 2015.

\bibitem{f3}
Cohl Furey.
\newblock ${SU(3)_C\times SU(2)_L \times U(1)_Y (\times U(1)_X)}$ as a symmetry
  of division algebraic ladder operators.
\newblock {\em Euro. Phys. J. C}, 78:375, 2018.

\bibitem{maithresh2019}
Maithresh Palemkota and Tejinder~P. Singh.
\newblock Proposal for a new quantum theory of gravity {III}: Equations for
  quantum gravity, and the origin of spontaneous localisation.
\newblock {\em Zeitschrift f\"ur Naturforschung A}, 75:143, 2019
  DOI:10.1515/zna-2019-0267 arXiv:1908.04309.

\bibitem{stcw}
Tejinder~P. Singh.
\newblock Space-time from collapse of the wave-function.
\newblock {\em Zeitschrift f\"ur Naturforschung A}, 74:147, arXiv:1809.03441,
  2019.

\end{thebibliography}

\end{document}